\documentclass[twocolumn]{aastex701}
\usepackage{amsmath,amssymb}
\usepackage{tikz}
\usetikzlibrary{arrows.meta,positioning,calc,fit,backgrounds}
 
\definecolor{ink}{RGB}{38,42,52}
\definecolor{hair}{RGB}{201,206,214}
\definecolor{mute}{RGB}{150,157,168}
\definecolor{shadebg}{RGB}{243,244,246}
\definecolor{arline}{RGB}{38,98,133}
\definecolor{arfill}{RGB}{234,242,247}
\definecolor{archip}{RGB}{215,231,241}
\definecolor{gbline}{RGB}{160,88,46}
\definecolor{gbfill}{RGB}{251,242,234}
\definecolor{gbchip}{RGB}{245,228,213}
\definecolor{resline}{RGB}{34,84,66}
\definecolor{resfill}{RGB}{233,242,237}
\definecolor{neuchip}{RGB}{228,231,236}

\usepackage{amsmath}

\begin{document}

\title{Probing AU-Scale Magnetic-Field Reversals in the Interstellar Medium with Pulsar Scintillation}

\author[orcid=0009-0000-0935-9245,sname='Jacob Yen']{Jacob Yen\begin{CJK}{UTF8}{bsmi}(顏丞宇)\end{CJK}}
\altaffiliation{Institute of Astronomy and Astrophysics, Academia Sinica}
\affiliation{Institute of Astronomy and Astrophysics, Academia Sinica}
\email[show]{jacobyen2001@gmail.com}  

\author[orcid=0000-0001-7888-3470,gname=Daniel, sname=Baker]{Daniel Baker} 
\affiliation{Institute of Astronomy and Astrophysics, Academia Sinica}
\email{dbaker@asiaa.sinica.edu.com}

\author[orcid=0000-0001-7931-0607,gname=Dongzi,sname=Li]{Dongzi Li}
\affiliation{Department of Astronomy, Tsing Hua University}
\email{dongzili@princeton.edu}

\author[orcid=0000-0003-2155-9578,sname=Ue-Li,gname=Pen]{Ue-Li, Pen}
\affiliation{Institute of Astronomy and Astrophysics, Academia Sinica}
\affiliation{Canadian Institute for Theoretical Astrophysics}
\affiliation{Canadian Institute for Advanced Research (MaRS Centre)}
\affiliation{Dunlap Institute for Astronomy and Astrophysics, University of Toronto}
\affiliation{Perimeter Institute of Theoretical Physics}
\email{pen@asiaa.sinica.edu.tw}

\begin{abstract}
AU-scale structures in the interstellar medium have become relevant across several areas of astrophysics, from ISM microphysics to pulsar timing arrays and cosmic-ray (CR) propagation. Current sheets formed at magnetic-field reversals have drawn particular attention because they can deflect the dominant GeV CR population through large angles. Pulsar scintillation, an outstanding probe of plasma density, resolves these structures — but not their magnetic configuration. Conventional Faraday rotation measures (RM) are only marginally sensitive at AU scales and are further compromised by ionospheric systematics. 

We propose a method that isolates birefringence-induced polarization---the signature of magnetic reversal---by applying phase retrieval to polarized data. Pulsar scintillation further resolves ray paths separated by $\sim 1\,\rm{AU}$, enabling differential RM measurements that suppress the ionospheric common mode while preserving the AU-scale magnetic signal. We demonstrate the method on an archival observation of PSR B0834+06. Phase retrieval resolves the 1-ms scattering feature into two branches in Doppler-delay space. Under the corrugated current-sheet interpretation, these branches correspond to ray paths sampling opposite sides of an AU-scale current sheet. We measure a branch-to-branch RM difference of $(-9.3\pm3.2)\times 10^{-3}\text{ rad m}^{-2}$  at 2.9$\sigma$ across a 10-MHz band. The signal is consistent with a field reversal of $|\Delta\langle B_\parallel \rangle| \simeq4.4\pm2.1\,\mu\rm G$, sufficient to sustain a long-lived sheet. With only six independent subbands, we present this as a proof of concept rather than a definitive detection. If it proves scalable, its full implementation could constrain cosmic-ray transport and inform models of propagation-induced noise in pulsar timing arrays.
\end{abstract}

\keywords{\uat{Interstellar Scintillation}{855} --- \uat{Interstellar Magnetic Fields}{845} --- \uat{Interstellar Scattering}{854} --- \uat{Radio Pulsars}{1353} --- \uat{Cosmic Ray}{329} --- \uat{Warm Ionized Medium}{1788}}




\section{Introduction} 


The flow of energy and matter within the interstellar medium (ISM) is mediated by cosmic rays (CRs), which play a critical role in modulating large-scale galactic dynamics \citep{CR_gal_evo_feedback_Zweibel_2017, gal_feedback_2023Ruszkowski}. However, certain small-scale structures of the ISM could drastically affect CR trajectories. Recent work demonstrates that localized small-scale magnetic-field reversals produce current sheets, causing large-angle deflection in CR propagation \citep{2023Kempski,2023Lemoine,2024Butsky}. This is in direct opposition to the conventional picture where CR diffusion is treated as globally smooth and nearly isotropic \citep{stochastic_1969Jokipii,eff_diffusion_2001Strong, stochastic_2002Schlickeiser}. Specifically, the dominant CR population, the GeV CRs \citep{GeV_2015_Grenier}, has a gyroradius comparable to astronomical-unit (AU) scales. Current research proposes that the ramification of the coinciding scales is that intermittent AU-scale current sheets could largely influence CR propagation \citep{2025Kempski}. In a general ISM warm ionized medium phase, the deviation of CR diffusion coefficients will lead to systematic overestimation of the inferred CR outflow efficiencies, with potential consequences for star-formation regulation and galaxy evolution. These considerations motivate the concrete characterization of AU-scale current sheets.

However, the magnetic field configuration in the diffuse ISM remains poorly characterized on AU-scales, both theoretically and observationally. In the classic MHD turbulence framework, an inertial range spans several orders of magnitude in spatial scale, with a Kolmogorov-like energy spectrum in the direction perpendicular to the local magnetic field \citep{Goldriech_Sridhar1995}. But as the scale decreases, the ISM becomes increasingly intermittent, fragmenting into structures such as current sheets \citep{2004Schekochihin, 2015Chandran, 2022Dong}. This corresponding dissipation scale remains loosely constrained \citep{2014Momferratos, 2024Lesaffre}. Observations, conversely, have strong evidence for AU-scale ISM plasma substructure, from pulsar scintillation \citep{Stinebring2001,2004Walker,2005Hill, 2006Cordes} and extreme scattering events \citep{Fiedler_1987, Romani_1987}. Pulsar scintillation arises from pulsar radiation scattering, followed by interference between multiple propagation paths. From the interference pattern, electron density fluctuations are inferred to be on the AU scale. Extreme scattering events (ESEs) are more drastic scatterings in which background-source brightness undergoes transient dimming due to refractive lenses crossing the line of sight. The refractive lens is typically estimated at AU scales as well. However, both phenomena inform us merely about density structures on AU scales; magnetic field configuration on such scales remains observationally unconstrained.

Traditionally, magnetic fields are probed using the Faraday rotation effect on polarized sources (such as pulsars), in which a magnetized plasma rotates the polarization angle of linearly polarized light. The rotation measure (RM), which quantifies the magnetic field strength, is defined with a density-weighted integral over distance. This distance, for a typical pulsar, spans hundreds of parsecs, so sensitivity to AU-scale magnetic structures is strongly diluted in the diffuse ISM. Long-term monitoring of a pulsar's proper motion could, in principle, sample RM variations across AU-scale transverse range \citep{RM_AU_2004Weisberg, RM_AU_2010Weisberg, RM_AU_2011Yan}. However, such measurements are limited to ionospheric Faraday rotation, which introduces a time-variable common-mode contribution \citep{iono_2013Sotomayor,iono_2019Porayko}. At the precision required to resolve AU-scale variations, the ionospheric term is difficult to separate from interstellar RM variation. These factors together have prevented direct observational constraints on AU-scale magnetic field configuration using conventional RM measurements.

\begin{figure}
    \centering
    \includegraphics[width=0.8\linewidth]{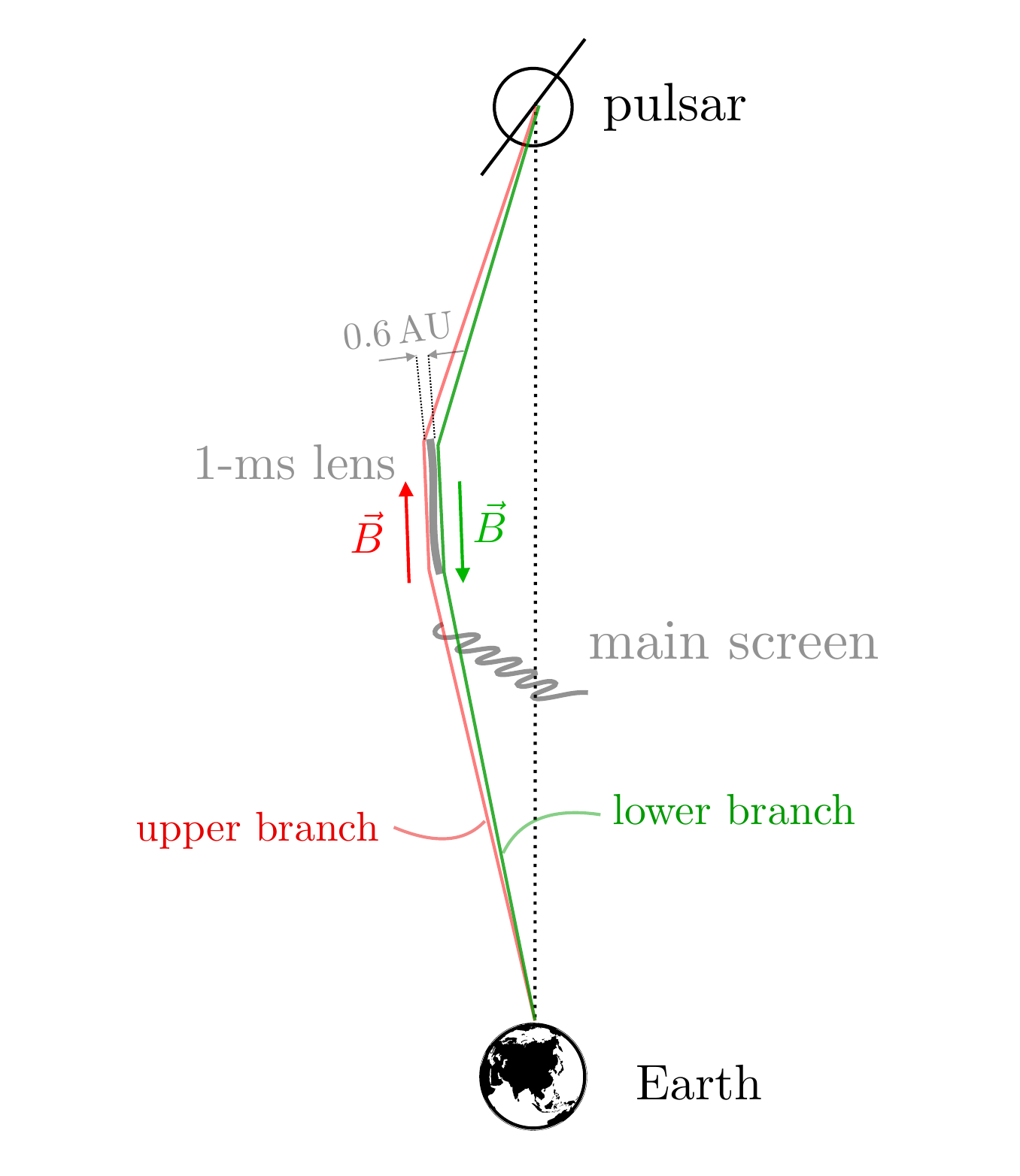}
    \caption{Schematic geometry of the double-lens configuration. Rays from PSR B0834+06 are first refracted by the secondary ``1-ms lens'' and subsequently by the main scattering screen. In the corrugated current-sheet framework, the main screen is modeled as a corrugated sheet, while the 1-ms lens is interpreted as a current sheet elongated toward the line of sight. A magnetic-field reversal is expected across the 1-ms sheet. The two resulting ray paths produce the observed upper and lower branches of the 1-ms feature in the conjugate wavefield, corresponding to rays that graze opposite faces of the sheet.}
    \label{fig:1sche}
\end{figure}

We probe the magnetic field configuration of an AU-scale foreground structure using interstellar scintillation to circumvent these limitations. Adopting our methodology, we probe the magnetic field configuration of a foreground structure toward PSR B0834+06, known as the ``1-ms feature" \citep{2010Brisken, 2022Baker}. In the corrugated current-sheet framework, the feature arises from a double-lensing geometry with a main screen and a strong secondary screen. We refer to the latter as the ``1-ms lens''. The 1-ms lens is modeled as a current sheet elongated along the line of sight with transverse size on AU scales \citep{Zhu_2023}. The 1-ms lens focuses two distinct ray paths from the pulsar emission, manifesting as two branches in the Fourier spectrum (upper and lower branches), and each branch encodes the magnetic field contribution from one side of the sheet (Figure \ref{fig:1sche}). The key to this method comes from leveraging the main screen to amplify the signal of the two lensed rays in AU separation. The mechanism will be expounded in Section 3. The morphology of the two rays and their manifestation in the data are illustrated in Figure \ref{fig:msf}. The corrugated current sheet picture predicts a magnetic field reversal as the mechanism to form the current sheet---the magnetic field should point in opposite directions on either side of the sheet. We therefore expect samples of consistent branch-to-branch RM difference across frequencies.

To achieve AU scale resolution, \textit{phase retrieval} must complement the standard intensity-based scintillation toolkit. Specifically, we implement the $\theta-\theta$ method \citep{2022Baker,Baker_2023} to perform a differential measurement of the left and right circular polarizations (LCP and RCP). While standard phase retrieval analysis on scintillation probes matter distributions, pulsar emission traversing magnetic-field reversals exhibits birefringence-induced polarization \citep{Macquart_2000}, unreachable by general phase retrieval algorithms in pursuit of scalar propagation phase. Our approach uniquely isolates these birefringence-induced modes and is furthermore generalizable to unpolarized background sources. Crucially, our method incorporates two additional downstream differencing operations: isolating the background line-of-sight RM, and subsequently calculating the spatial RM difference between the upper and lower branches. This dual-subtraction achieves a simultaneous cancellation of the common-mode ionospheric contribution, consequently pushing the RM resolution by approximately two orders of magnitude below the baseline ionospheric uncertainty \citep{iono_2019Porayko}. If proven operational, this method can provide valuable observational input for models of CR transport and constrain propagation effects in pulsar timing arrays. 

We will first present our data in Section 2 to visualize scintillation, and then briefly cover pulsar scintillation theory in Section 3, i.e., explain how the scintillation pattern in the previous section arises. Section 4 explains our method in full detail, and in Section 5 we demonstrate the result on our archival data as a proof-of-concept, and end with discussion in Section 6.

\section{Data}

The data of PSR B0834+06 we analyze are archival data, observed 20 years ago in November 2005 as part of a global VLBI project described in \citet{2010Brisken}. They used the largest telescopes at that time: Arecibo (AR), Green Bank Telescope (GBT), Jodrell Bank (JB), and the tied-array Westerbork (WB), recording both circular polarizations. We adopt data only from AR and GB as they are the highest signal-to-noise set. The observation spans a total bandwidth of 32 MHz over $310.5-342.5$ MHz, composed of four 8-MHz bands with a frequency resolution of 244 Hz and gated integration time of 1.25 s. The total on-source time is 5700 s, while the observation spans $\sim 2$ hr. As the power of scintillation predominantly lies in Stokes parameters I and V, scintillation observations commonly record Stokes I and V only. Since the data were not intended for polarimetry to begin with, cross-hand spectra (Stokes Q and U) are therefore absent. Raw complex phases encode unstable pulsar emission and instrumental phases in addition to that induced by interstellar scintillation. Quadratic quantities carry cleaner information about scintillation; hence, the complex electric fields are not stored.

We follow the data reduction details according to \citet{Simard_2019}; we adopt the dynamic spectrum $|E_L|^2$ and $|E_R|^2$ of both AR and GB, as well as the visibility cross spectrum $E_L^\text{AR}(E_L^\text{GB})^*$ and $E_R^\text{AR}(E_R^\text{GB})^*$. For subsequent analysis, we averaged the dynamic spectrum and the visibility cross-spectrum over a 5 s integration.

\begin{figure}
    \centering
    \includegraphics[width=\linewidth]{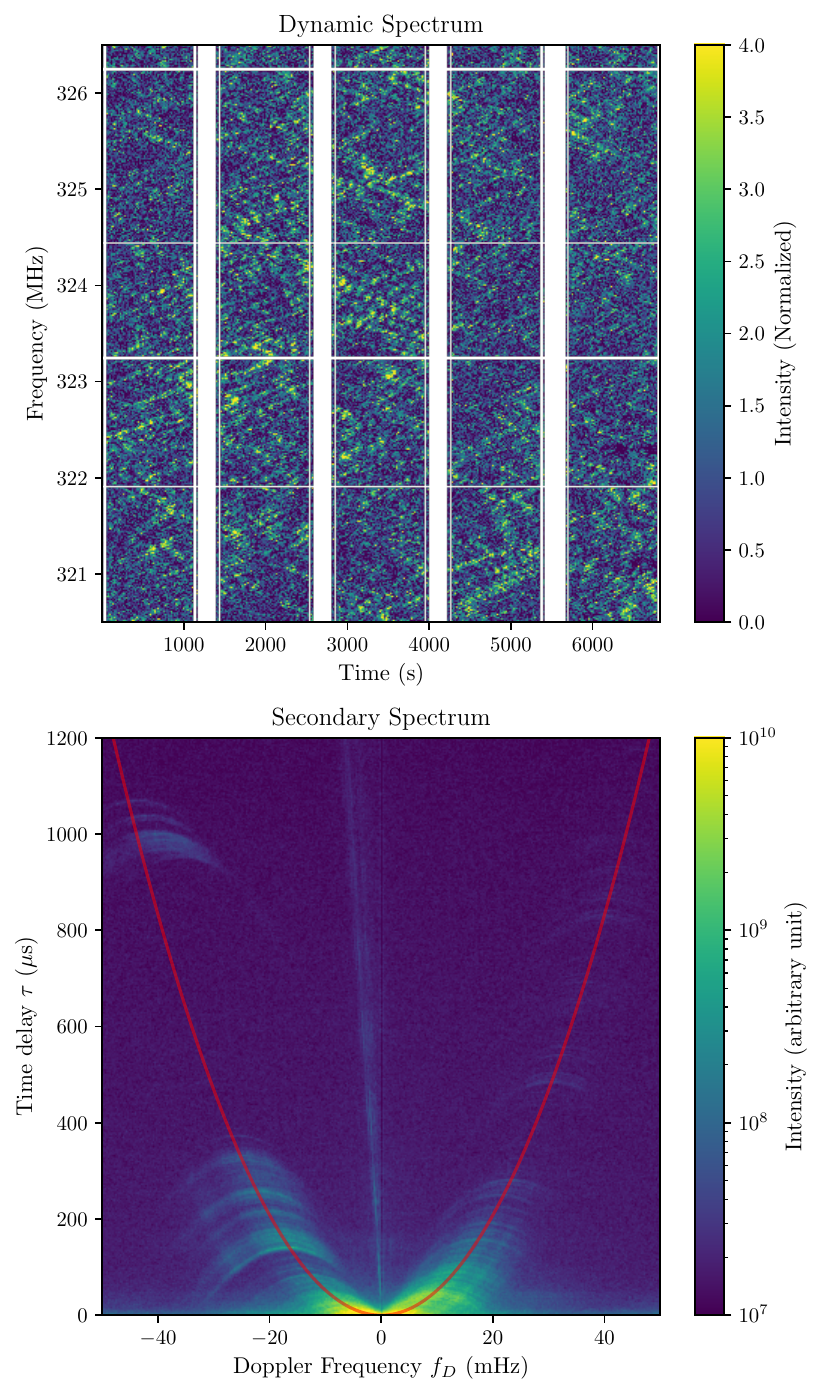}
    \caption{(Top) A portion of the dynamic spectrum of PSR B0834+06, showing characteristic criss-cross intensity modulations produced by interference between multiple scattered propagation paths. (Bottom) The corresponding secondary spectrum, defined as the power spectrum of the dynamic spectrum, exhibits a prominent scintillation arc accompanied by inverted arclets. Note that the island power at $\tau\sim1000\,\mu \rm s$ will later be referred to as the 1-ms feature.}
    \label{fig:2dyn}
\end{figure}

\section{Pulsar Scintillation}

Pulsar scintillation is a low-frequency effect due to scattering of radio waves into multiple paths, forming images on a localized screen, and subsequently multipath propagation generates interference patterns observed on Earth. The characteristic intensity modulations in the dynamic spectrum showing criss-cross patterns arise from interference between multiple scattered propagation paths produced by localized plasma density inhomogeneities. The dynamic spectrum $I(\nu,t)=|E(\nu,t)|^2$, the modulus squared of the complex electric field, is shown on the top panel of Figure \ref{fig:2dyn} (from Arecibo).

\subsection{Scintillation arcs}

Moreover, the interference pattern exhibits a highly organized structure---scintillation arcs---in the secondary spectrum, the power spectrum of the dynamic spectrum $S(\tau,f_D) = |\tilde I(\tau,f_D)|^2$, where $\tau$ is the time delay conjugate to frequency and $f_D$ is the Doppler frequency conjugate to time. In this representation, the power concentrates around a well-defined parabolic arc $\tau=\eta f_D^2$ for some curvature $\eta$, first discovered by \citet{Stinebring2001} for several pulsars. A more comprehensive survey of scintillation arcs can be found in \citet{Stinebring_2022}.

Following the discovery of scintillation arcs, several theoretical works studying secondary spectra \citep{2004Walker,2006Cordes} suggest that the production of such arc morphology requires two conditions: (1) the scattering happens in a rather confined region relative to the entire line of sight \citep{2005Hill}; we later refer to the region as a (scattering) screen. (2) The scattered images should be highly anisotropic in theory, i.e., close to one-dimensional. The anisotropy of images was confirmed by the novel scintillation-based VLBI astrometry of \citet{2010Brisken}; PSR B0834+06 shows images effectively one-dimensional (See their Figure 2 and 5), which is generally true for a large fraction of pulsars.

One might note that the inverted arclets are ordered instead of random. The scintillation arc can be understood as an interference pattern generated by images, where radio flux incidentally converges on a scattering screen. These images serve as secondary sources that re-emit the electromagnetic wave from the pulsar. We prescribe the position of an image by the angular position vector $\boldsymbol\theta_i$. The constructive interference in the secondary spectrum is concentrated specifically at $(\tau,f_D)$ satisfying

\begin{equation}\label{eq:taufd2d}
    \begin{aligned}
        \tau &= f_\nu = \frac{d_\text{eff}}{2c}(|\boldsymbol\theta_j|^2-|\boldsymbol\theta_k|^2),\\
        f_D &= f_t = \frac{1}{\lambda}(\boldsymbol\theta_j-\boldsymbol\theta_k)\cdot \mathbf{v}_\text{eff}
    \end{aligned}
\end{equation}

If the images were spread two-dimensionally, the secondary spectrum will result in a diffuse continuum filling the interior of the parabola. For one-dimensional scattering that results in a well-defined, cleaner morphology of a thin arc without inverted arclets (such as some in \citet{Stinebring_2022}) arises when the line-of-sight image at $\boldsymbol\theta=0$ dominates; the arc arises from the interference between the line-of-sight image and the scattered images. Here, the above trivially reduces to $\tau=\frac{d_\text{eff}}{2c}\theta^2,\ f_D = \frac{v_\parallel}{\lambda}\theta$, and therefore the quadratic relation $\tau = \eta f_{\small D}^{\ 2}$, with curvature
\begin{equation}
\eta = \dfrac{d_\text{eff}\lambda^2}{2cv_\parallel^2}
\end{equation}
This frequency dependence of arc curvature $\eta\propto \lambda^2\propto\nu^{-2}$ will occur later. Otherwise, when the scattered images are sufficiently bright compared to the line-of-sight image, the main arc is accompanied by some inverted arclets with the same curvature $\eta$ as a general case for Eq. \ref{eq:taufd2d}; the inverted arclets result from the interference among scattered images themselves.

\subsection{The 1-ms feature}

Also, there can be two or more scattering screens between the pulsar and Earth. In the upper left region of the secondary spectrum in Figure \ref{fig:2dyn}, there is an island of power, named the ``1-ms feature" due to its position on the spectrum. It is a resulting feature of scattering from an offset secondary screen from the line of sight, which is known as the 1-ms lens from the introduction. Figures 2 and 5 in \citet{2010Brisken} have the 1-ms lens as a blob away from the linear image configuration.

It is inferred that the 1-ms lens is a strong lens with a locally high electron column density $N_e$, which \citet{Zhu_2023} suggests will cause an extreme scattering event (ESE) if crossing the line of sight. This reasoning is nearly model-independent, invoking only one mechanism from geometrical optics: The point of largest gradient $dN_e/dx$ will focus the radio emission locally to form a \textit{caustic}---a brightness peak. Two caustics are consistent with one $N_e$ peak and could be approximated by the Gaussian plasma lens. This is the standard model for ESE without any assumptions about the constituting material of the ESE lens; see Figure 2 from \citet{1998Clegg} for a great visualization of an ESE light curve resulting from a refractive Gaussian electron column density distribution. The manifestation of two caustics corresponds to the upper and lower branches in our Figure \ref{fig:msf}. Note that \citet{Zhu_2023} used phase retrieval to reproduce the two branches, allowing their conclusion that the 1-ms lens is a strong lens. Figure \ref{fig:msf} is a reproduction of former works rather than a discovery.

\subsection{The corrugated current sheet framework}

The conclusive evidence of one-dimensional scintillation geometry from \citet{2010Brisken} has triggered many to propose their models to physically explain this phenomenon. In fact, \citet{2010Brisken} in their paper already proposed two preliminary models to account for their discovery. \citep{Gwinn_2019} also proposed a one-dimensional plasma structure, which he himself referred to as the noodle model.

\citet{PenLevin_2014} developed a geometric model based on corrugated current sheets, assuming the least contrived physical mechanisms to reproduce the 1D feature. Qualitatively, they succeeded in reproducing the scintillation arc and the 1D scattering image of PSR B0834+06. Subsequently, \citet{Simard_2018} realized the idea into a quantitative model specifying the geometry of plasma lensing. Recently, \citet{Jow_2024} completed the big picture by modeling the 1-ms feature with \textit{catastrophe theory} from mathematics. They not only reproduced the deconvolved morphology of the 1-ms feature, but also provided a coherent explanation for the low- and high-frequency ESE light curve. The latter is exceptionally challenging in that many ESE models could reproduce the characteristic low-frequency light curve, as in \citet{1998Clegg}, but struggle at high frequency. 

The universal framework uniting scintillation and ESEs classified the corrugated sheet as an A2 fold catastrophe, while ESE lenses such as the 1-ms lens are an A3 cusp catastrophe. Though mathematically elegant, it could be perceived as sophisticated rather than simple. Contrarily, simplicity is the central philosophy of this framework that echoes the rationale of \citet{PenLevin_2014} to assume minimally on turbulence or reconnection-driven mechanisms. Instead, the entire thread only hinges on the geometry of the plasma density distribution. 

The 1-ms lens modeled as an A3 lens, to put it simply, is approximated by a two-dimensional current sheet viewed at grazing incidence \citep{Zhu_2023, Jow_2024}. A current sheet, according to Ampere's law, necessitates the existence of a magnetic-field reversal on its opposite sides. A current sheet in pressure equilibrium requires itself to increase in electron density $n_e$ compared to the surrounding. Therefore, a 2D flat current sheet viewed edge-on observes a column density increase by definition $N_e = \int_L n_e dl$ for a current sheet of characteristic length $L$. This minimally explains the Gaussian $N_e$ profile of \citet{1998Clegg}. 

Although this corrugated current sheet framework provides strong explanations for scintillation and ESE, the reproduced phenomena are primarily due to plasma density structure. If the 1-ms lens is a current sheet, then we are motivated to find out its magnetic signal---the magnetic field reversal on opposite sides.

\section{Methods}

To measure branch-to-branch RM offset, we first apply phase retrieval to recover the LCP and RCP phases discarded in the recording, obtaining two wavefields per station. A Fourier transform of these wavefields resolves the 1-ms feature into an upper and a lower branch, which remain separable in the six lowest frequency subbands. Each branch carries its own differential rotation measure ($\Delta\rm{RM}$), isolated by derotating the line-of-sight phase to zero. The difference between the two branches then yields six independent samples of $\Delta(\Delta\rm{RM})$, from which we determine the offset and its uncertainty. This offset is convertible to a difference in the line-of-sight magnetic field through the corrugated current sheet framework.

\begin{figure}
\centering
\includegraphics[width=\linewidth]{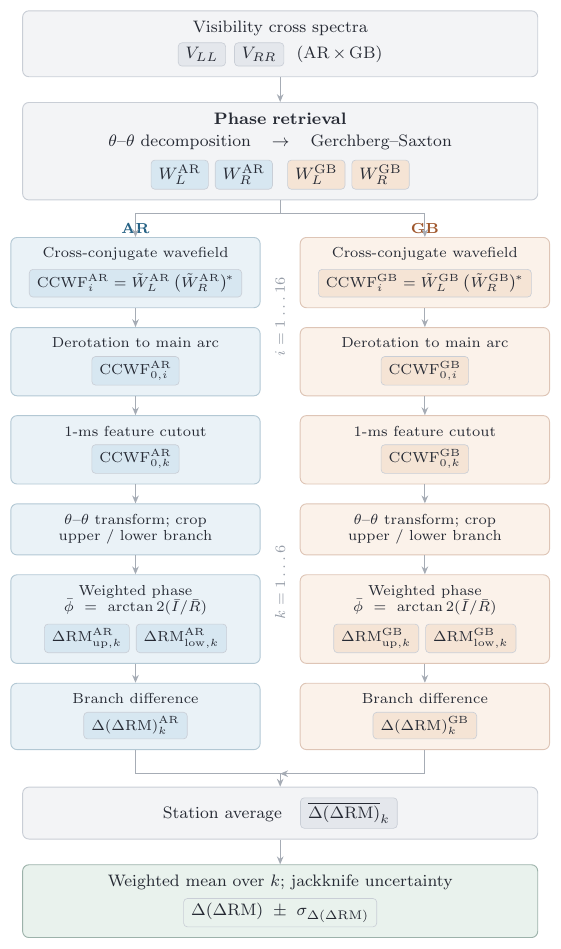}
\caption{Analysis pipeline. Phase retrieval is applied to the
visibility cross-spectra to recover one wavefield per station and
polarization. The cross-conjugate wavefield is formed per station,
derotated to the main arc in each of 16 subbands, and the 1-ms feature
is isolated in the six subbands where the two branches remain resolved.
A representative phase is extracted from each branch box and converted
to $\Delta$RM; the branch difference is formed per station, averaged
over stations, and combined across subbands.}
\label{fig:pipeline}
\end{figure}

Figure \ref{fig:pipeline} summarizes the pipeline; Table \ref{tab:quantities} defines the quantities used throughout.

\begin{deluxetable*}{llll}
\tablecaption{Quantities used in the phase-retrieval analysis.
\label{tab:quantities}}
\tablehead{
\colhead{Quantity} & \colhead{Symbol} & \colhead{Definition} &
\colhead{Domain}
}
\startdata
Raw electric field & $E(\nu,t)$ & Complex: Phase from pulsar, instrumental, and scintillation  & $(\nu,t)$ \\
Dynamic spectrum & $I(\nu,t)$ & Real: $|E(\nu,t)|^{2}$; discarding the phases & $(\nu,t)$ \\
Visibility cross spectrum & $V(\nu,t)$ & Complex: $E^{\rm AR}(E^{\rm GB})^{*}$, $V_{LL}$, $V_{RR}$ per polarization & $(\nu,t)$ \\
\hline
Wavefield & $W(\nu,t)$ & Complex: Phase only from scintillation  & $(\nu,t)$ \\
Conjugate wavefield & $\tilde W(\tau,f_D)$ & Complex: 2D FT of $W(\nu,t)$ & $(\tau,f_D)$ \\
Conjugate (dynamic) spectrum & $\tilde I(\tau,f_D)$ & Complex: 2D FT of $I(\nu,t)$; $\tilde I= \tilde W \otimes \tilde W^{*}$ & $(\tau,f_D)$ \\
Conjugate visibility spectrum & $\tilde V(\tau,f_D)$ & Complex: 2D FT of $V(\nu,t)$ & $(\tau,f_D)$ \\
Secondary spectrum & $S(\tau,f_D)$ & Real: $|\tilde I(\tau,f_D)|^{2}$, for visualizing the complex conjugate wavefield & $(\tau,f_D)$ \\
\hline
Cross-conjugate wavefield & CCWF & Complex: $\tilde W_L \tilde W_R^{*} \propto e^{i(\phi_L-\phi_R)}$, where $\phi_{L,R}$ defined in $(\tau,f_D)$ & $(\tau,f_D)$ \\
\enddata
\tablecomments{$E_L$, $E_R$ per polarization; Cross-correlation visibilities $V_{RL},\ V_{LR}$ are not recorded. $\otimes$ denotes autocorrelation. Single-dish phase
retrieval applies an eigenvector decomposition to $\tilde I$ to extract
the dominant mode; VLBI visibility requires an SVD of $\tilde V$ to
extract one eigenmode per station.}
\end{deluxetable*}

\subsection{Phase Retrieval}
The dynamic spectrum, by construction, removes the phases, which motivates the phase retrieval problem to recover a complex wavefield from intensity data. Note that the purpose of phase retrieval, in the context of interstellar scintillation, is to only reconstruct the phase arising from scintillation, without the chaotic phase from the pulsar intrinsic emission and instrumental phases. 

Several methods are proposed to perform phase retrieval on pulsar scintillation data, including iterative stacking images/arcs \citep{2005WalkerStinebring, Pen_2014}, cyclic spectroscopy \citep{2013Walker,2024Turner}, and the $\theta-\theta$ method \citep{2022Baker, Baker_2023} . 

The $\theta-\theta$ method generalizes \citet{Pen_2014} and is particularly powerful for one-dimensional scattering geometries. We opt for this method since our data matches this geometric condition. It turned a mathematically difficult problem into a linear algebra problem with existence and uniqueness statements. After the $\theta-\theta$ reconstruction, an additional step---the Gerchberg-Saxton algorithm---is able to recover the 1-ms feature where the previous method fell short.

\subsubsection{$\theta-\theta$ Technique}
The first step extracts the main arc, and the latter deconvolves the conjugate dynamic spectrum ($\tilde I(\tau,f_D)$, whose modulus square is the secondary spectrum) with that main arc to produce the conjugate wavefield $\tilde W(\tau,f_D)$, as $\tilde I = \tilde W \otimes\tilde W^*$. 

First step, the decomposition hinges on the $\theta-\theta$ transformation \citep{Sprenger_2020}, which maps parabolic arcs from the conjugate dynamic spectrum $\tilde I(\tau,f_D)$ into linear features in the $\theta-\theta$ space, via the coordinate transformation

\begin{equation}\label{eq:ttt}
\begin{aligned}
    \theta_1 &= \frac{1}{2}\left(\frac{\tau}{\eta f_D} + f_D\right) \\ 
    \theta_2 &= \frac{1}{2}\left(\frac{\tau}{\eta f_D} - f_D\right) ,
\end{aligned}
\end{equation}
turning it into a linear algebra problem. The method captures the dominant eigenmode of the linear feature in $\theta-\theta$ space as a result, because a one-dimensional geometry corresponds to a rank-1 matrix structure in $\theta-\theta$ representation mathematically. The output is a rank-1 approximation of the underlying wavefield, a primitive conjugate wavefield $\tilde W (\tau,f_D)$, encoding only the main arc phase information, with a phase gauge freedom that is intrinsic to the algorithm. The $\theta-\theta$ method applies to both single-dish and VLBI data. For single-dish data, we apply eigenvector decomposition to extract the eigenmode from the conjugate dynamic spectrum $\tilde I(\tau,f_D)$. For VLBI visibility spectra, we use SVD to extract two eigenmodes for individual stations from the conjugate visibility spectrum $\tilde V(\tau,f_D)$.

Since the wavefield is not coherent over the entire bandwidth and time scale, and the frequency-dependence of curvature that $\theta-\theta$ transform inputs, we do not construct a $\theta-\theta$ map from the entire spectrum. Instead, the spectrum is chunked into overlapping frequency-time chunks to doubly tile the spectrum. Upon obtaining phase retrieval solutions in every chunk, they are then pieced together by aligning the phase gauge in overlapping regions. Our selected chunk size is $\sim$ 125 kHz $\times$ 600s, large enough for sufficient S/N for the $\theta-\theta$ reconstruction, but local enough to achieve approximate coherence. This process is called ``mosaicking." However, the solution of the phase map after mosaicking is still underdetermined by an overall constant phase, intrinsic to the $\theta-\theta$ method. 

\subsubsection{The Gerchberg-Saxton algorithm}

\begin{figure}
    \centering
    \includegraphics[width=\linewidth]{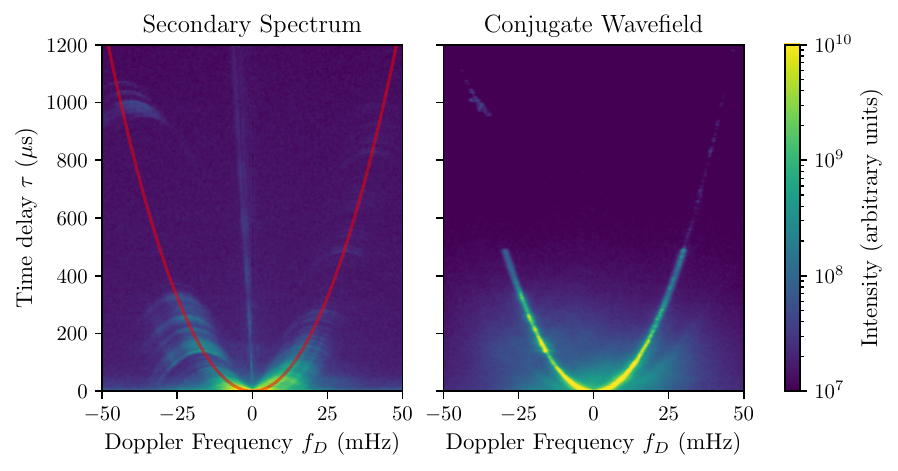}
    \caption{(Left) The secondary spectrum from observation. (Right) The absolute value of the complex-valued conjugate wavefield, produced by the phase retrieval $\theta-\theta$ method. Note that each point is associated with a retrieved phase.}
    \label{fig:deconv}
\end{figure}

Then, the Gerchberg-Saxton algorithm iteratively enforces consistency with the dynamic spectrum on the primitive wavefield and the causality constraint in the conjugate space until the wavefield converges, with the presence of the deconvolved 1-ms feature. The real-valued visualization of the input and output is shown in Figure \ref{fig:deconv}. The secondary spectrum (modulus squared of the conjugate dynamic spectrum) and the modulus of the conjugate wavefield. For further details, see \citet{2022Baker, Baker_2023} for reference.

In our analysis, we adopt the visibility cross-spectrum from the AR-GB VLBI for the $\theta-\theta$ phase-retrieval method, as single-dish yields a null result. The advantage of visibility is that it suppresses single-dish systematics and increases phase stability, in turn providing higher S/N. The  $\theta-\theta$ method receives the conjugate visibility spectrum $\tilde V$ as the input, and yields one reconstructed wavefield per station, while each wavefield is underdetermined by an arbitrary phase gauge.  The universality of the constant phase over the full bandwidth breaks after the GS algorithm, which necessitates the band-by-band derotation.

\subsection{Cross-Conjugate Wavefield and Phase Derotation}

Upon obtaining the wavefield with $\theta-\theta$ phase retrieval for LCP and RCP, the conjugate wavefield---defined as the 2D Fourier transform of the wavefield---for both polarizations is generated ($\tilde W_L,\tilde W_R$). The phase difference between the two wavefields (which informs the Faraday rotation angle) is acquired by multiplying one conjugate wavefield by the complex conjugate of the other; the result is referred to as the \textit{cross-conjugate wavefield}
\begin{equation}\label{eq:diff1}
    \text{CCWF} \equiv \tilde W_L\tilde W_R^*\propto e^{i(\phi_L-\phi_R)}. 
\end{equation}
In the following analysis, the CCWF is also divided into 16 frequency subbands (2 MHz subbands), motivated by the evolution of scintillation arc curvatures over frequency (both the main arc and the 1-ms feature evolve $\propto 1/\nu^2$). 

To analyze the phase of the 1-ms feature in the CCWF across these 16 subbands, the phase gauge must be fixed to a consistent reference, as the $\theta-\theta$ method introduces a consistent phase gauge to wavefields. To fix this reference, we apply independent derotation to each 2-MHz subband of CCWF by setting the mean phase of the dominant power---the main arc (the high-S/N feature arising from line-of-sight interference) to zero, synonymously, derotating the mean phase of the dominant main arc to zero. This is numerically attained by optimizing the derotation angle to minimize the imaginary component of the main-arc power while maximizing its real component. This operation fixes the line-of-sight phase as the reference (Eq. \ref{convert}), so that the residual phase in the 1-ms feature corresponds to the $\Delta\rm{RM}$. The resulting derotation angles for Arecibo and Green Bank are shown in Figure \ref{fig:derot} (More details are in the Appendix).

\begin{figure}
    \centering
    \includegraphics[width=0.9\linewidth]{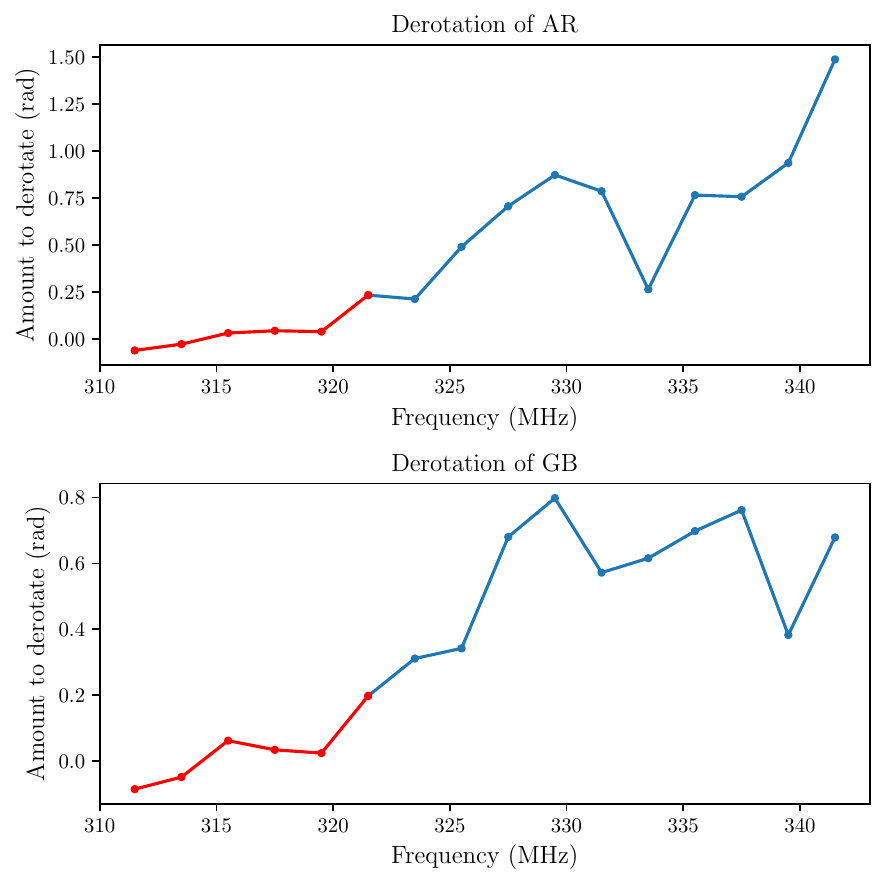}
    \caption{The amount of derotation (in radians) for 16 frequency bands, that is, setting the average phase of the main arc in the CCWF to be 0. The phase of the 1-ms feature is measured relative to this zero-point. The first six bands attract our focus since the upper and lower branches in the 1-ms feature remain separated; therefore, we mark them in red. }
    \label{fig:derot}
\end{figure}

\subsection{$\Delta$RM Measurement Procedure}

The Faraday rotation produces a wavelength-dependent phase difference $\phi$ between LCP and RCP wave, with the relation $\frac{1}{2}\phi=\lambda^2\rm{RM}$ where the rotation measure (RM) is defined as
\begin{equation}
    \text{RM}  = 0.81\int n_e B_\parallel d\ell\ .
\end{equation}
Here, $n_e$ is the electron density in $\text{cm}^{-3}$, $B_\parallel$ is the magnetic field parallel to the ray path direction $d\vec\ell$ in microgauss ($\mu\rm G$), and $d\ell$ is in parsec (pc).
After derotation, the derotated phase $\Delta\phi=(\phi-\phi_\text{los})$ converts to our observable $\Delta\rm{RM}$ by 
\begin{equation}\label{convert}
    \frac{1}{2}\Delta\phi=\lambda^2\Delta\rm{RM},
\end{equation}
where $\Delta\text{RM}=\text{RM}-\text{RM}_\text{los}$ denotes the RM of any ray path relative to the line-of-sight RM. 

The 1-ms feature is curved in conjugate space; to analyze its phase, we map it into the $\theta-\theta$ space where parabolae (and inverted parabolae) are straightened into horizontal (and vertical) lines. The upper and lower features, as two horizontal line segments, are isolated by cropping out two boxes around them (see Figure \ref{fig:result}, right). The box sizes are chosen to fully encompass the coherent width of each branch.

A representative phase of a box enclosing a finite-width line is obtained by a weighted mean over the box in $\theta-\theta$ space. For each branch enclosed in a box, we obtain the weighted phase with the following steps: (i) We sum the power over the $\theta_1$ axis. To represent the signal distribution as a function of $\theta_2$, we fit a Gaussian profile to define the signal weight. (ii) Using this Gaussian weight, we do a normalized weighted sum of the complex array $C_{ij}=R_{ij}+iI_{ij}$ over $\theta_2$ and obtain a one-dimensional power projection $C_i=R_i+iI_i$ as a function of $\theta_1$. (iii) We compute the weighted mean of the real and imaginary parts $(\bar R,\bar I)$ using $|R_i|$ as the weight, and obtain the phase of each box with $\bar \phi=\arctan2(\bar I/\bar R)$. The weighting scheme emphasizes pixels with dominant power and reduces sensitivity to noise-dominated regions. This measured weighted phase $\bar\phi$ corresponds to the derotated phase $\Delta\phi$, which can be converted into $\Delta$RM using \eqref{convert}. 

We also propose a robust error estimate of this measurement. Since the conjugate space has undergone a Fourier transform, and $\theta-\theta$ transformation is also non-linear, the sampling of noise is non-trivial. The estimation of noise for measurements of each box is as follows. We first randomly select 8 noisy regions (where the main arc does not pass) in the conjugate wavefield, and also apply the same $\theta-\theta$ transform with the same curvature. We go through the same weighting process to find a weighted mean phase as a sample of the noise, yielding 8 noise samples for both real and imaginary parts. We then take the standard deviation $\sigma_{\bar R},\sigma_{\bar I}$ and propagate the noise through the function $\arctan(\bar I/\bar R)$ as
\begin{equation}
    \sigma_{\bar\phi} = \frac{|\bar R|}{\bar R^2+\bar I^2}\sqrt{\frac{\bar I^2}{\bar R^2}\sigma_{\bar R}^2+\sigma_{\bar I}^2}\ .
\end{equation}
The phase uncertainty is obtained by first-order propagation of independent fluctuations in $\sigma_{\bar R}$ and $\sigma_{\bar I}$; the derivation is provided in the Appendix.

To obtain $\Delta(\Delta\rm{RM})$ as a function of frequency, we take the difference of $\Delta\rm{RM}$ between the two branches at each frequency band. The frequency-independent offset is then estimated using an inverse-variance-weighted mean of the $\Delta(\Delta\rm{RM})$ values, with the associated uncertainty given by the standard inverse-variance error. This procedure yields a statistically weighted estimate of the branch-to-branch RM offset.

The pipeline should yield $\Delta(\Delta\text{RM})$ over 6 bands. Due to the systematics dominating in single stations, the branches are not separate until we combine the results of two stations by averaging. The result is presented as follows.

\section{Results}

\subsection{The deconvolved 1-ms feature}

\begin{figure}
    \centering
    \includegraphics[width=\linewidth]{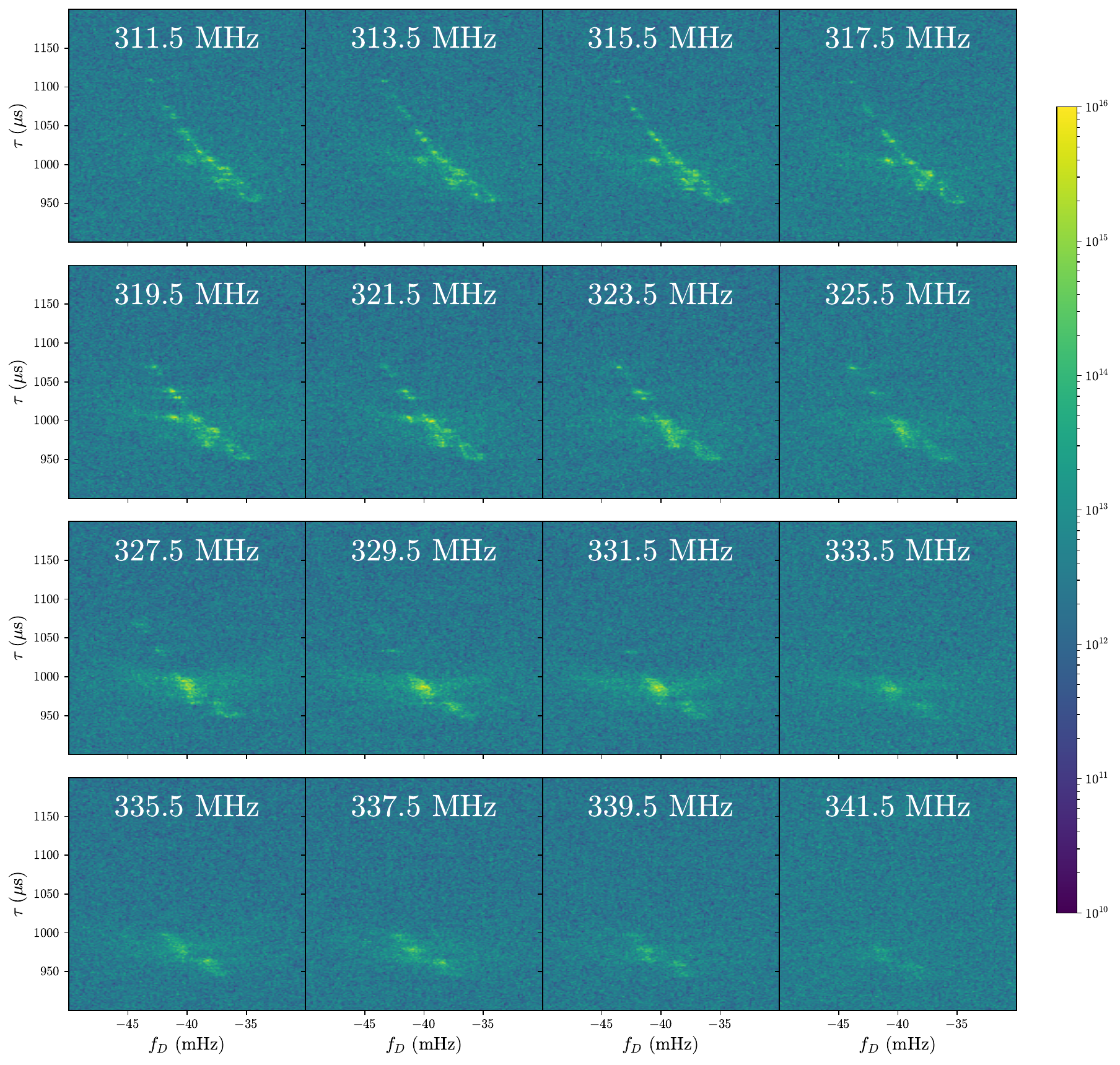}
    \caption{Frequency evolution of the 1-ms feature across a 32 MHz-wide band, divided into 2 MHz subbands. Two separated arc-like branches are visible in lower-frequency bands and progressively merge and weaken at higher frequencies. The plotted signal is the real part of the derotated cross-conjugate wavefield (CCWF); the derotation procedure and CCWF definition are described in Eq. \ref{eq:diff1}.}
    \label{fig:msf}
\end{figure}

Figure~\ref{fig:msf} shows the 1-ms feature recovered in the conjugate
wavefield following $\theta-\theta$ phase retrieval and Gerchberg--Saxton
deconvolution. Where the secondary spectrum contains a single island of power
near $\tau\approx1000\,\mu$s (Figure ~\ref{fig:deconv}), the deconvolved wavefield
resolves the feature into two distinct branches, each with a curved, half-arc
morphology.

To characterise the frequency evolution of the feature, we divide the 32~MHz
band into 16 subbands of 2~MHz. The two branches are clearly separated in the six lowest subbands ($311.5$--$321.5$~MHz). Their angular separation decreases monotonically with increasing frequency, consistent with the $\nu^{-2}$ scaling of the arc curvature expected from scintillation theory  \citep{2003Hill,2006Cordes}. Above $323.5$~MHz the separation falls below the resolution of the reconstruction and the branches merge into a single feature, accompanied by a decrease in amplitude toward higher frequencies. All measurements of the branch-to-branch difference below are therefore restricted to the six lowest subbands.

\begin{figure*}
    \centering
    \includegraphics[width=\textwidth]{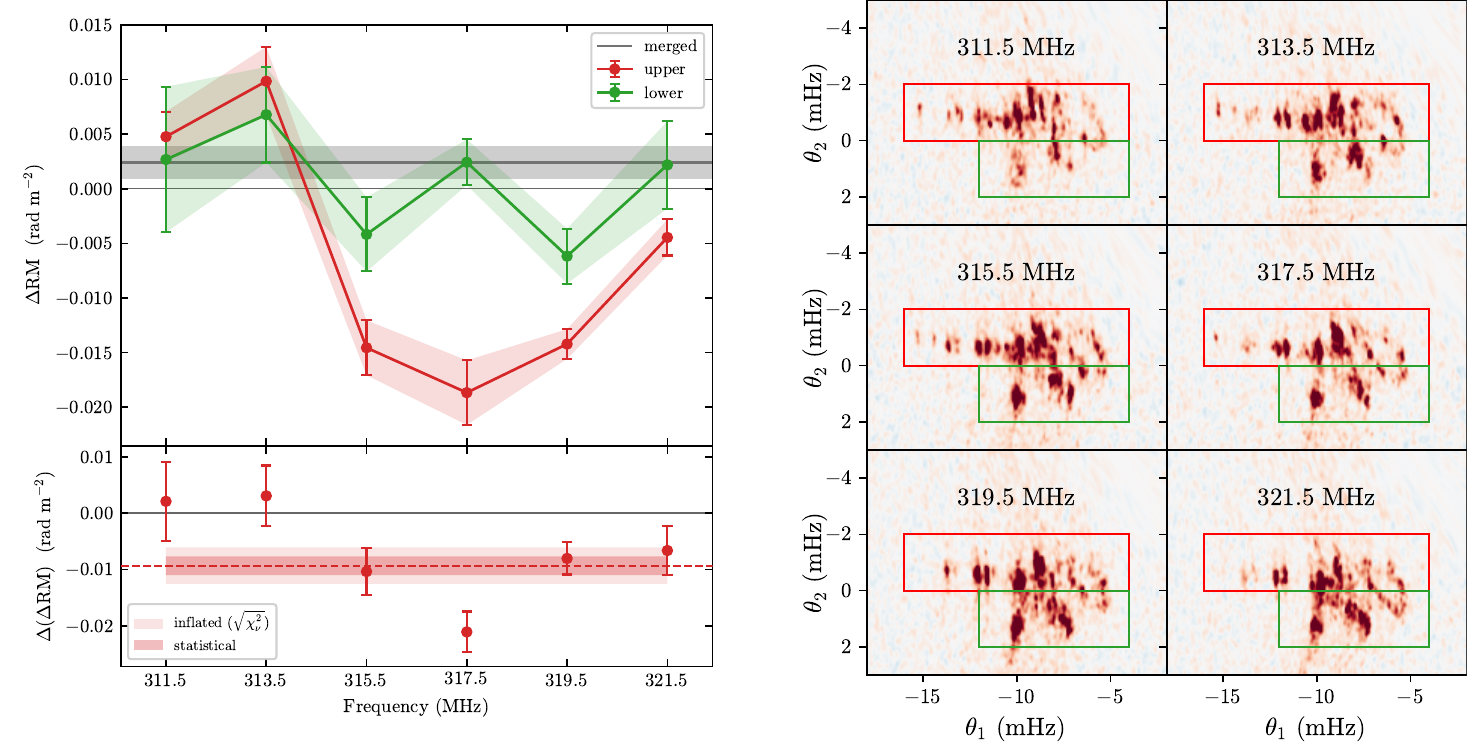}
    \caption{(Left) Primary measurement. \emph{Top}: $\Delta$RM for the upper (red) and lower (green) branches across the six frequency bands in which the two branches remain resolved in $\theta-\theta$ space; shaded regions give the per-band uncertainties. The grey band marks the merged-feature measurement above $323.5$~MHz, $(2.4\pm1.5)\times10^{-3}$~rad~m$^{-2}$, where the branches are no longer separable. \emph{Bottom}: the branch-to-branch difference $\Delta(\Delta\rm{RM})$, with the weighted mean as a dashed line. The dark band is the inverse-variance uncertainty; the light band is that uncertainty rescaled by $\sqrt{\chi^2_\nu}$ to reproduce the observed band-to-band scatter ($\chi^2_\nu = 3.83$, $\nu=5$), which we adopt. The rescaled result is $(-9.3\pm3.2)\times10^{-3}$~rad~m$^{-2}$, a $2.9\sigma$ offset; under the sheet interpretation, this corresponds to $|\Delta\langle B_\parallel\rangle| = 4.4\pm2.1\,\mu$G. (Right) The 1-ms feature in $\theta-\theta$ space for the same six bands (real part of the derotated CCWF, as in Figure ~\ref{fig:msf}), showing the rectangular regions used to isolate the upper (red) and lower (green) branches.}
    \label{fig:result}
\end{figure*}

\subsection{Differential RM of the individual branches}

We measure $\Delta\rm{RM}$ for each branch in the six subbands where they remain separable. In each subband, the branches map to horizontal line segments in $\theta-\theta$ space (Figure ~\ref{fig:result}, right); rectangular regions
enclosing each segment are used to extract a representative phase by the
weighting procedure of Section~4.3, and the phase is converted to
$\Delta\rm{RM}$ via Eq.~\eqref{convert}. Per-band values are listed in
Table~\ref{tab:ddrm} and plotted in Figure ~\ref{fig:result} (left, top).

Both branches show band-to-band variation in excess of their per-band
uncertainties. A constant fit to $\Delta\rm{RM}_{\rm upper}$ gives
$\chi^2_\nu = 21.7$ ($\nu=5$), and to $\Delta\rm{RM}_{\rm lower}$ gives
$\chi^2_\nu = 2.3$. Such variation is expected: the line-of-sight zero point is
determined independently in every subband through main-arc derotation
(Section~4.2), so each branch measurement carries an independent calibration
offset in addition to measurement noise. Neither branch constitutes a significant measurement on its own, and we do not interpret them
individually.

\begin{deluxetable}{lccc}
\tablecaption{Per-band rotation measure differences for the two branches of the
1-ms feature, and the branch-to-branch offset.\label{tab:ddrm}}
\tablehead{
\colhead{$\nu$ (MHz)} &
\colhead{$\Delta\rm{RM}_{upper}$} &
\colhead{$\Delta\rm{RM}_{lower}$} &
\colhead{$\Delta(\Delta\rm{RM})$} \\
\colhead{} &
\colhead{($10^{-3}\,\rm{rad\,m^{-2}}$)} &
\colhead{($10^{-3}\,\rm{rad\,m^{-2}}$)} &
\colhead{($10^{-3}\,\rm{rad\,m^{-2}}$)}
}
\startdata
$311.5$ & $  4.76 \pm 2.29$ & $  2.67 \pm 6.66$ & $  2.09 \pm 7.04$ \\
$313.5$ & $  9.83 \pm 3.15$ & $  6.77 \pm 4.37$ & $  3.06 \pm 5.39$ \\
$315.5$ & $-14.55 \pm 2.51$ & $ -4.18 \pm 3.39$ & $-10.37 \pm 4.21$ \\
$317.5$ & $-18.66 \pm 2.95$ & $  2.42 \pm 2.09$ & $-21.08 \pm 3.62$ \\
$319.5$ & $-14.21 \pm 1.36$ & $ -6.17 \pm 2.51$ & $ -8.04 \pm 2.85$ \\
$321.5$ & $ -4.46 \pm 1.65$ & $  2.18 \pm 4.00$ & $ -6.64 \pm 4.33$ \\
\hline
Weighted mean            & $-8.06$ & $-0.39$ & $-9.33$ \\
\quad $\sigma$ (inverse variance) & $0.82$  & $1.28$  & $1.66$  \\
\quad $\sigma$ (jackknife)  & $4.16$  & $2.56$  & $3.24$  \\
$\chi^2_\nu$ (constant fit) & $21.7$  & $2.3$   & $3.8$   \\
Significance             & $1.9\sigma$ & $0.2\sigma$ & $2.9\sigma$ \\
\enddata
\tablecomments{Band centres are spaced by $2$~MHz.  Per-band uncertainties are the noise-sample estimates of Eq.~(4); $\Delta(\Delta\rm{RM}) = \Delta\rm{RM}_{upper}-\Delta\rm{RM}_{lower}$ with errors added in quadrature.
The constant fit has $\nu=5$ degrees of freedom.  Because $\chi^2_\nu>1$ in all three columns, the inverse-variance errors are underestimated and we adopt the delete-one jackknife error, quoted as the significance; for $\Delta(\Delta\rm{RM})$ this coincides with the inverse-variance error inflated by $\sqrt{\chi^2_\nu}$ ($3.24\times10^{-3}\,\rm{rad\,m^{-2}}$).  The unweighted scatter error $s/\sqrt{N}=3.63\times10^{-3}\,\rm{rad\,m^{-2}}$ gives a consistent, slightly more conservative value.  At $N=6$ the fractional uncertainty on $\sigma$ itself is $\sim32\%$.}
\end{deluxetable}

\begin{deluxetable}{lccc}
\tablecaption{Estimators for the branch-to-branch offset.\label{tab:est}}
\tablehead{\colhead{Estimator} & \colhead{$\Delta(\Delta\rm{RM})$} &
\colhead{$\sigma$} & \colhead{Significance}\\
\colhead{} & \multicolumn{2}{c}{($10^{-3}\,\rm{rad\,m^{-2}}$)} & \colhead{}}
\startdata
Inverse-variance weighted        & $-9.33$ & $1.66$ & $5.6\sigma$ \\
\quad $\times\sqrt{\chi^2_\nu}$  & $-9.33$ & $3.24$ & $2.9\sigma$ \\
Jackknife (delete-one)           & $-9.33$ & $3.24$ & $2.9\sigma$ \\
Unweighted, $s/\sqrt{N}$         & $-6.83$ & $3.63$ & $1.9\sigma$ \\
\enddata
\tablecomments{$N=6$. The inverse-variance error is rejected as the
per-band errors are underestimated ($\chi^2_\nu=3.8$).  We adopt the
jackknife, which coincides with the inflated inverse-variance error and
makes no Gaussian assumption on the per-band errors.  At $N=6$ the
fractional uncertainty on $\sigma$ itself is $\sim32\%$.}
\end{deluxetable}

\subsection{The branch-to-branch RM difference}

We define the branch-to-branch difference
\begin{equation}\label{eq:ddrm}
\Delta(\Delta\mathrm{RM})\equiv\Delta\mathrm{RM}_{\rm upper}
-\Delta\mathrm{RM}_{\rm lower},
\end{equation}
evaluated within each subband, so that the derotation reference is common to
both branches and cancels in the difference. Per-band values appear in the
fourth column of Table~\ref{tab:ddrm} and in the lower panel of
Figure ~\ref{fig:result}.

An inverse-variance weighted mean of the six values gives
$-9.33\times10^{-3}$~rad~m$^{-2}$ with a formal uncertainty of
$1.66\times10^{-3}$~rad~m$^{-2}$. The corresponding constant fit has
$\chi^2_\nu \approx 3.83$ ($\nu=5$), indicating that the per-band uncertainties
underestimate the true dispersion of the measurements by a factor of
approximately two. We therefore do not adopt the formal uncertainty.

Rescaling the formal uncertainties so that the constant fit reproduces the observed band-to-band scatter gives $3.24\times10^{-3}$~rad~m$^{-2}$. A delete-one jackknife, which does not assume the individual per-band uncertainties beyond their use as relative weights, gives $3.24\times10^{-3}$~rad~m$^{-2}$, agreeing to within three significant figures. An unweighted mean of the six bands with the standard error of the mean gives $(-6.8\pm3.6)\times10^{-3}$~rad~m$^{-2}$, consistent with the weighted result and marginally more conservative. We adopt the jackknife uncertainty and report
\begin{equation}\label{eq:main}
\Delta(\Delta\mathrm{RM}) = (-9.3\pm3.2)\times10^{-3}\ \mathrm{rad\ m^{-2}},
\end{equation}
a $2.9\sigma$ offset from zero. With six bands, the fractional uncertainty on the uncertainty itself is $\approx30\%$, so the distinction between the rescaled and unweighted estimates is not meaningful, and we do not treat the precise significance level as informative beyond indicating a marginal measurement.

\subsection{Frequency dependence and individual bands}

Under the interpretation adopted here, $\Delta(\Delta\rm{RM})$ is a property of the intervening medium and should be independent of observing frequency. We test this by fitting polynomial terms in frequency across the six subbands. Using the rescaled uncertainties of Section~5.3, adding a linear term reduces $\chi^2$ by $0.40$ for one additional degree of freedom, and adding a quadratic term reduces it by $2.84$ for two ($p=0.24$). Neither term is required by the data, and we adopt the constant model.

The $317.5$~MHz band carries the largest residual and contributes $55\%$ of the total unrescaled $\chi^2$. However, this deviation is tolerable statistically; because the raw errors are initially underestimated by a factor of $\sqrt{\chi_\nu^2}\approx 1.96$, this maximum residual corresponds to merely $1.7\sigma$ as the true observed scatter after scaling. We therefore retain all six subbands. Crucially, this band possesses the second smallest intrinsic uncertainty; it carries $21\%$ of the total statistical weight. The fact that its residual remains bounded at $1.7\sigma$ despite this heavy weighting demonstrates it is a physically driven feature rather than an artifact of unmodeled noise. The sensitivity of the amplitude to a single band is a limitation of the present dataset, and is the principal reason we regard the measurement
as a demonstration of the method rather than a precise determination.

\subsection{Merged bands}

Above $323.5$~MHz the two branches are unresolved, and no branch-to-branch difference can be formed. In these ten subbands we measure a single $\Delta\rm{RM}$ for the merged feature, obtaining an inverse-variance weighted mean of $(2.4\pm1.5)\times10^{-3}$~rad~m$^{-2}$, consistent with zero. This range is shaded grey in Figure ~\ref{fig:result} (left). We report this as a consistency check on the derotation procedure at high frequency; because the merged feature is a different observable from the branch difference, it does not directly constrain systematics affecting Eq.~\eqref{eq:ddrm}. 

\subsection{Conversion to a magnetic-field difference}

Under the sheet geometry of \citet{Zhu_2023}, the branch-to-branch RM
difference converts to a difference in the density-weighted parallel magnetic field via 

\begin{equation}\label{bmeas}
\Delta\langle B_\parallel\rangle =
\frac{\Delta(\Delta\mathrm{RM})}{0.81\,\frac{dN_e}{dx}\,w},
\end{equation}

where $dN_e/dx$ is the transverse electron column density gradient, and $w$ is the physical width of the 1-ms lens; the formula is derived in Appendix~A. Three assumptions enter this conversion. First, the two rays before impinging on the 1-ms lens take approximately the same paths, so that all foreground and background contributions cancel in the difference. Second, the electron column encountered by the two rays at the edge of the lens is assumed equal, so that the residual difference is attributed entirely to $B_\parallel$; we have not measured the branch-to-branch column difference independently. Third, the lens parameters are taken from a specific model of the feature.

Adopting $dN_e/dx \simeq 900\ \rm{cm}^{-3}$, inferred from the bending angle of the image \citep{1998Clegg, Zhu_2023}, and $w = 0.6\pm0.2$~AU, derived from the magnification and impact parameter, the measured offset corresponds to 

\begin{equation}
|\Delta\langle B_\parallel\rangle| \simeq 4.4\pm2.1\ \mu\rm{G},
\end{equation}

where the uncertainty combines the fractional errors on
$\Delta(\Delta\rm{RM})$ and on $w$ in quadrature and does not include any uncertainty on $dN_e/dx$. If the magnetic field is exactly antisymmetric about the sheet center, this difference corresponds to a characteristic magnitude of $\approx2.2\ \mu$G on either side.

Since $\Delta N_e \simeq (dN_e/dx)\,w$, Eq.~\eqref{bmeas} may be written $\Delta\langle B_\parallel\rangle = \Delta(\Delta\text{RM})/(0.81\,\Delta N_e)$, and the fractional uncertainties combine as 

\begin{equation}
\left(\frac{\sigma_B}{B}\right)^{2}=
\left(\frac{\sigma_{\Delta(\Delta\rm{RM})}}{\Delta(\Delta\rm{RM})}\right)^{2}
+\left(\frac{\sigma_{\Delta N_e}}{\Delta N_e}\right)^{2}.
\end{equation}

We adopt $\sigma_{\Delta N_e}/\Delta N_e = \sigma_w/w = 33\%$ from the width quoted by \citet{Zhu_2023}, and do not include a separate uncertainty on $dN_e/dx$, which is derived from the same lens model and is not independently constrained.

\section{Discussion}

We demonstrate the method on the 2005 archival data from \citet{2010Brisken}, obtaining a marginal $2.9\sigma$ signal, as a proof of concept rather than a definitive detection. We have shown that the result is consistent with the theory of interstellar scintillation and the corrugated current sheet picture.

\subsection{Limitations and improvements}

Although the methodology may indicate the presence of an AU-scale magnetic-field reversal, the result is limited by its statistical significance (2.9$\sigma$) and by its reliance on a single pulsar and a single observing epoch. The ubiquity and temporal stability of this phenomenon, therefore, remain to be established through future observations. 

We also acknowledge that rank-1 reconstruction of the wavefield serves as a first-order approximation; future work incorporating higher ranks into the reconstruction may provide fruitful results.

Across $311.5-321.5$ MHz, the change in $\lambda^2$ is $\Delta\lambda^2=0.057\,\mathrm{m}^2$. A systematic term proportional to $\lambda^2$ would therefore vary by only $\sim6.5\%$ across the observed bandwidth and is difficult to distinguish from the present data. More fundamentally, the leakage can happen between RCP and LCP. Rigorous polarimetry records all correlation modes: RR, LL, RL, LR between two stations, while only RR and LL are accessible to us. RL and LR correspond to generating Stokes Q and U that constrain the leakage.

\subsection{Confirmation of framework}

Scintillation theory is demonstrated by the deconvolved 1-ms feature curvature scaling as $\nu^{-2}$. Concretely, the apex of their extended parabolas $(\tau_0,f_{D,0})$ is centered somewhere at $f_{D,0}<0,\ 500<\tau_0<1000$, offset from the origin because the image is not on the line of sight. Parametrizing both branches by the form $\tau-\tau_0 = \eta_\text{1ms}(f_D-f_{D,0})^2$, two branches close in at higher frequencies due to $\eta_\text{1ms}\propto\nu^{-2}$.
The observed morphology of the two branches  (Figure \ref{fig:msf}) is consistent with the geometry of a convergent (underdense) lens modeled by the corrugated current-sheet framework, matching the left-side image in Figure 9 from \citet{Jow_2024}.

\subsection{Observation bottleneck advancement}
Furthermore, the ability to observe RM differences between extremely fine angular separations, and thus AU transverse scale for the ISM, marks a unique highlight of this method. This is made possible by combining scintillation with sharp frequency resolution of this uniquely well-suited data. By measuring the representative phase of two branches with the described weighting scheme, the derotated phase of the cross-conjugate wavefield is isomorphic to a $\Delta$RM map, the RM of every point relative to the line-of-sight RM, explicitly, 
\begin{equation}
\begin{aligned}
    \Delta\text{RM}(\boldsymbol{\theta}) &= (\text{RM}_\text{true}(\boldsymbol{\theta})+\text{RM}_\text{ion}(\boldsymbol{\theta}))\\
    &\quad -(\text{RM}^\text{los}_\text{true} +\text{RM}^\text{los}_\text{ion}). \\
    & = \text{RM}(\boldsymbol{\theta}) - \text{RM}^\text{los}
\end{aligned}
\end{equation}
The derotation achieves the purpose of cancelling the common-mode ionospheric contribution, as the representative ray paths of each point are just separated from the line of sight by orders of meters ($\mathcal O(\theta)\sim$ mas), effectively the same sky area $\text{RM}_\text{ion}(\boldsymbol{\theta})\approx\text{RM}^\text{los}_\text{ion}$. This concept had been proposed to better model the ionospheric term, but not for the purpose of measuring high spatial resolution \citep{iono_2013Sotomayor}. Measuring $\Delta\text{RM}$ down to $\sim 0.01\ \text{rad m}^{-2}$ level is sharp in units of RM resolution, compared to the general ionospheric uncertainty of $\sim 1\text{ rad m}^{-2}$ currently, although comparison based on different metrics, the former is transversed $\Delta$RM, the latter is along one sightline.

\subsection{ISM structure}

This branch-to-branch difference in Faraday rotation corresponds to a magnetic-field reversal with amplitude $4.4 \pm 2.1\,\mu\text{G}$ based on the corrugated current-sheet picture. The inferred field-reversal amplitude is physically plausible in the context of the plasma beta, defined as $\beta=P_\text{th}/P_B$ (with thermal pressure $P_\text{th}=nk_BT$ and magnetic pressure $P_B=B^2/8\pi$ in cgs units). Adopting representative parameters for the warm ionized medium ($P_\text{tot}/k_B\approx3000 \ \rm K\text{ cm}^{-3}$ in relatively dense diffuse-ISM regions; see Appendix for details), we obtain $\beta\approx 1$. This value indicates that thermal and magnetic pressures are of comparable magnitude, consistent with pressure equilibrium, hence allowing this structure to be long-lived.

Small-scale current sheets also arise naturally in MHD simulations of energy cascades, which predict magnetic coherent structures such as current sheets to form ubiquitously on small (sub-pc) scales \citep{2023Fielding,2024Ntormousi}. Extrapolation of these results down to AU-scales, however, remains resolution-limited. Taken together, these comparisons show qualitative agreement with observation and simulation, and comparable amplitude scaling among theories.

On the theory of ESE lenses (like the 1-ms screen), historical alternative models include photo-ionized skins of a self-gravitating neutral dense cloud \citep{1998Walker_Wardle} and axisymmetric cylindrical Gaussian plasma lenses \citep{2018Dong}. The self-gravitating cloud scenario is increasingly disfavored, because the implied total mass in such clouds would be comparable to the Galactic mass budget and may require additional nonstandard assumptions (e.g., dark-matter-like populations). 

Subsequent ESE model developments have therefore emphasized less exotic plasma-lens scenarios involving localized density enhancements. However, axisymmetric cylindrical filament models face challenges in pressure confinement and geometric stability. In this context, the sheet-based interpretation naturally explains both the lensing morphology and the RM signal. Although not uniquely required, the corrugated current sheet is therefore favored by the present observations.

\subsection{Pulsar timing arrays}
Intermittent sheets associated with ESEs are also relevant to pulsar timing array (PTA) experiments, because when an ESE lens lies close to the line of sight, it can produce multiple images with distinct geometric time delays, analogous to scintillation, thereby perturbing the pulse arrival time. The bulk scattering delay induced by scintillation is $\sim 50-200$ ns \citep{Main2020}, comparable to the $\sim100$ ns timing precision targeted for detection of a stochastic gravitational-wave background \citep{noise_floor_2005Jenet}. Given an ESE occurrence rate of $\sim1\%$ \citep{1994Fiedler}, systematic multi-epoch and multi-sightline surveys of small-scale magnetic structures will be important for assessing their prevalence and temporal stability. By then, incorporating physically motivated ESE-lens models into scattering-delay corrections may help reduce timing residuals and improve PTA sensitivity to the gravitational-wave background.

\subsection{CR propagation and future observations}

Our first measurement provides evidence for an opposite magnetic field across the face of an AU-scale elongated current sheet, as described in \citet{Zhu_2023}. This structure contributes to strong scattering of the dominant GeV CR population, where cosmic ray physics and galaxy evolution in general are at stake.

This geometry is also discussed in \citet{2025Kempski} on CR propagation, our measured $\Delta$RM magnitude of $\sim0.01\text{ rad m}^{-2}$ matches their predicted RM fluctuations, with explicit scaling of $\Delta$RM in their Eq. (24) as 
\begin{equation}
    \Delta\text{RM} \sim 0.05\ \text{rad m}^{-2}\frac{\Delta n_e}{0.1\,\text{cm}^{-3}}\frac{\ell}{10^4\text{AU}}\frac{B_\parallel}{10\,\mu\rm G}.
\end{equation}
The physical width $w=0.6\pm 0.2$ AU is measured by \cite{Zhu_2023}, which also gave a rough aspect ratio $\ell/w\sim10^3-10^4$; this scales the $\Delta$RM to roughly down to $O(0.001-0.01)$, which our measurement of $\Delta(\Delta\text{RM})=0.0093\pm0.0032$ rad m$^{-2}$ falls in this regime.

Detecting additional ``1-ms feature''-like events will require high-cadence monitoring together with high-resolution interferometric follow-up. Such capabilities are becoming available at several radio facilities, including wide-field pulsar monitoring arrays and long-baseline interferometric networks operating in target-of-opportunity modes. To perform precise polarimetry, all four Stokes parameters are required. The Q and U parameters are required to characterize the polarization leakage, but are not accessible in this work.

For a general pulsar scintillometry observation, the observation setup is highly dependent on the system's geometry. Resolutions in $f_D$ and $\tau$ are set by observing duration $1/T$ and total bandwidth $1/B$, while the extent of the Doppler-delay map $f_\text{D,max}$ and $\tau_\text{max}$ is set by integration time $1/2\delta t$ and channel width $1/2\delta\nu$. Also, the $\lambda^2$ degeneracy is increasingly separable at lower frequencies, as stated in the limitations. Therefore, broader frequency coverage at lower frequencies will provide greater leverage on breaking the degeneracy.

\section{Conclusion}

This work aims to detect AU-scale magnetic field reversals associated with current sheets, which are of interest across diverse areas. The AU-scale matter distribution was accessible to the pulsar scintillation toolkit, but the magnetic signal remains unattainable. We find that such magnetic reversals would produce birefringence-induced polarization, which, unfortunately, requires $\Delta\phi=\phi_R-\phi_L$, a quantity that scintillation observations discard in practice.

We therefore propose our method involving several differential measurements to isolate the birefringent term from pulsar scintillation data. First, we reconstruct the phase of individual hands with phase retrieval \citep{Baker_2023}, and convert the retrieved phase into RM via CCWF. To calibrate for the phase gauge introduced by $\theta-\theta$ retrieval, we perform derotation of the CCWF phase to set the line-of-sight component to zero, equivalent to getting $\Delta\rm{RM}$. Lastly, taking the difference between the features of interest, in our case the two branches of the 1-ms feature, we get $\Delta(\Delta\text{RM})$ that together cancels the uncertainty introduced by the second difference as a common mode. The signal is convertible to magnetic field reversal amplitude via Eq. \ref{bmeas}. 

The \citet{2010Brisken} data we adopt, observed in 2005, were carefully designed to capture high-time-delay features and remain the benchmark data after 21 years. The 1-ms feature was deconvolved into two distinct branches at only a 10-MHz-wide bandwidth, corresponding to 6 data points in our realization. We measure a $(-9.3\pm3.2)\times 10^{-3}\text{ rad m}^{-2}$ signal with a modest 2.9$\sigma$ significance, suggestive of a signature consistent with the magnetic field reversal, but we position this result as a proof-of-concept demonstration rather than claiming a detection. Future observations are necessary to claim a robust detection of AU-scale magnetic-field reversal, requiring multiple epochs for repeatability, and making them rigorous for further scientific usage requires more sightlines to be explored.

\begin{acknowledgments}
We thank the members of the Toronto Scintillometry Group for helpful discussions and guidance on this project, and in particular Marten H. van Kerkwijk for insightful feedback. We are grateful to Walter Brisken for providing the VLBI data used in this study, to Daniel Rearden for sharing the original \texttt{scintools} code. We also thank Dylan Jow and Dan Stinebring for valuable theoretical discussions. We acknowledge the Institute of Astronomy and Astrophysics, Academia Sinica (ASIAA) ,for providing high-performance computing and data storage resources that significantly contributed to this work. GPT-5 and Claude Opus were used for text polishing. This work was supported in part by the Natural Sciences and Engineering Research Council of Canada (NSERC) under grants RGPIN-2019-06770, ALLRP-586559-23, and RGPIN-2025-06396, the Canadian Institute for Advanced Research (CIFAR), the Ontario Research Fund-Research Excellence (ORF-RE, 72074697), and AMD AI Quantum Astro.
\end{acknowledgments}

\begin{contribution}

U.-L.P. and D.B. designed the research; J.Y. and D.B. performed the research; J.Y. analyzed data; D.L. joined the theoretical discussion; J.Y. wrote the paper.


\end{contribution}

\bibliography{sample701}
\bibliographystyle{aasjournalv7}



\appendix

\section{Derivation of magnetic field conversion}

In magnetized plasma, the linear polarization of pulsar emission will be rotated by an angle of  $\Delta\chi$ due to the Faraday rotation effect, as two circular polarizations travel at different phase velocities. The angle of rotation $\Delta\chi$ scales with wavelength squared with 
\begin{equation}
    \Delta\chi=\text{RM}\ \lambda^2
\end{equation}
where RM is derived to be
\begin{equation}
    \text{RM} = \frac{e^3}{2\pi(m_ec^2)^2}\int_\text{path} n_eB_\parallel d\ell \simeq 0.81\int_\text{path} n_eB_\parallel d\ell
\end{equation}
in cgs units, with units $n_e\ [\text{cm}^{-3}]$, $B_\parallel\ [\mu\rm G]$, $d\ell\  [\text{pc}]$.
Another relevant quantity is the dispersion measure (DM), also known as the electron column density ($N_e$)
\begin{equation}
    \text{DM}=N_e = \int_\text{path} n_e d\ell
\end{equation}
From here, the mean parallel magnetic field weighted by electron density is 
\begin{equation}
    \langle B_\parallel \rangle = \frac{\int_\text{path}n_eB_\parallel d\ell}{\int_\text{path}n_e d\ell}= \frac{\text{RM}}{\text{0.81 DM}}
\end{equation}
If we consider the difference in RM between a ray path $P$ and the line of sight (LOS),
\begin{equation}
    \Delta\text{RM} = 0.81\left(\int_{P}n_eB_\parallel d\ell - \int_\text{los}n_eB_\parallel d\ell\right)
\end{equation}
If we split the sightline into 3 parts: from the pulsar to the lens, lens: inside the 1-ms lens, out: after traversing the lens, and primed for the path traversing the lens. We have
\begin{equation}
    \Delta\text{RM} = (\text{RM}_P^\text{in}+\text{RM}_P^\text{1ms}+\text{RM}_P^\text{out})-(\text{RM}_\text{los}^\text{in}+\text{RM}_\text{los}^\text{no lens}+\text{RM}_\text{los}^\text{out})
\end{equation}
while the two paths are just slightly separate in space, the ``in'' and ``out'' pass through roughly the same environment ($\text{RM}_P^\text{in}\approx\text{RM}_\text{los}^\text{in}$, $\text{RM}_P^\text{out}\approx\text{RM}_\text{los}^\text{out}$), the difference lies in ``lens'' and ``no lens''. Therefore, 
\begin{align}
    \Delta\text{RM} &\simeq \text{RM}_P^\text{1ms}-\text{RM}_\text{los}^\text{no lens}=0.81\left(\int_\text{P, 1ms}n_eB_\parallel d\ell-\int_{\substack{\text{los,} \\ \text{no lens}}}n_eB_\parallel d\ell\right)\\
    &=0.81(\text{DM}_P^\text{1ms}\langle B_\parallel\rangle_P^\text{1ms}-\text{DM}_\text{los}^\text{no lens}\langle B_\parallel\rangle_\text{los}^\text{no lens})
\end{align}
Comparing the upper and lower branches corresponding to the two rays grazing opposite sides of the elongated sheet, this yields
\begin{equation}
    \Delta(\Delta\text{RM})=0.81(\text{DM}_\text{upper}^\text{ 1ms}\langle B_\parallel\rangle_\text{upper}^\text{1ms}-\text{DM}_\text{lower}^\text{1ms}\langle B_\parallel\rangle_\text{lower}^\text{1ms})
\end{equation}
and given that the boundary is symmetric on the two faces, we get the same DM (or $N_e$) at the edge of the 1-ms lens, giving
\begin{equation}
    \Delta\langle B_\parallel\rangle=\frac{\Delta(\Delta\text{RM})}{0.81\ N_e^\text{1ms}}  
\end{equation}
And we can estimate the DM (or the electron column density) within the length of the lens by leveraging the result of the maximum column density gradient $dN_e/dx \simeq 900\text{ cm}^{-3}$ and the physical width of the current sheet to be $w=0.6$ AU from \cite{Zhu_2023}, the maximum difference in column density is in orders of 
\begin{equation}
    \Delta N_e =|N_e^\text{lens}-N_e^\text{edge}|\simeq \frac{dN_e}{dx}w.
\end{equation}
Given that our lens is underdense, the edge has a greater column density ($N_e^\text{lens}\ll N_e^\text{edge}$), which implies $\Delta N_e\simeq N_e^\text{edge}$, yielding
\begin{equation}
    \Delta\langle B_\parallel\rangle=\frac{\Delta(\Delta\text{RM})}{0.81\ \frac{dN_e}{dx}w}  \ .
\end{equation}
Our result gives
\begin{equation}
    |\Delta\langle B_\parallel\rangle|\simeq 4.4\ \mu\text G\left(\frac{\Delta(\Delta\text{RM})}{0.0094\text{ rad m}^{-2}}\right)\left(\frac{dN_e/dx}{900\text{ cm}^{-3}}\right)^{-1}\left(\frac{w}{0.6\text{ AU}}\right)^{-1}.
\end{equation}

\subsection{The plasma beta}
The plasma beta is defined as the ratio of thermal and magnetic pressure, $\beta=P_\text{gas}/P_B$, where $P_\text{gas}=nkT$ and $P_B=B^2/8\pi$ expressed in cgs units, $k = 1.38\times 10^{-16}\text{ dyne/K}$. 
Given a canonical sum of the two $(P_B+P_\text{gas})/k\simeq3000\text{ K cm}^{-3}$, the magnetic pressure term is $P_B/k=1395.5\text{ K cm}^{-3}$ (for $B=2.2\ \mu\rm G$), yielding $P_\text{gas}/k=1604.5$. Thus, the plasma beta $\beta=1604.5/1395.5\approx1.1$

\section{Phase Retrieval}

To recover the wavefield $W(\nu,t)$ with phases, we perform an eigenvector decomposition and model the $\theta-\theta$ spectrum by the largest eigenmode $\mu_i$, whose outer product with itself yields the $\theta-\theta$ spectrum (a rank-1 matrix),

\begin{equation}
\tilde I_{ij}(\theta_1,\theta_2)=\mu_i(\theta_1)\mu_j^*(\theta_2). 
\end{equation}

Next, if we tile the x-axis with the recovered eigenvector and inverse $\theta-\theta$ transform back to conjugate space $(f_D,\tau)$, we get the conjugate wavefield $\tilde W(\tau,f_D)$ with power concentrated on only the main parabola. The inverse FFT of it gives the wavefield $W(\nu,t)$.

Since the curvature of the arc varies with frequency, we divide the dynamic spectrum with a bandwidth of 32 MHz and a time span of 6825 s (131072 frequency $\times$ 1366 time channels) into overlapping chunks of $512\times128$ pixels ($125 \text{ kHz} \times 600\ \rm s $), such that these chunks doubly tile the full spectrum. As the $\theta-\theta$ method requires a curvature, we fit a curvature for each chunk and obtain a wavefield solution for each chunk. We perform this aware that coherence of the wavefield can only hold within a decorrelation scale. Then, they are assembled via mosaicking---a process that aligns overlapping tiled regions by phase. The practical reason is that $\theta-\theta$ solution is underdetermined up to a constant phase; therefore, to stitch the chunks together, we need to derotate the phase offset between two adjacent chunks. A Hann window function is applied as the weighting for the constant-overlap-add (COLA) to align the phase in the wavefield.

\section{Derotation}

Our goal is to analyze the phase of the 1-ms feature in the CCWF across these 16 subbands; therefore, the phase gauge must be fixed to a reference. The $\theta-\theta$ method introduces a global phase gauge; to calibrate the phase offset to a zero point, we need to ``derotate'' the amount of offset in individual frequency bands. However, the amount of phase offset measured from each subband is not a constant. Since we are weight-mapping a structured complex field (CCWF) into a single scalar, phase weighting is dominated by bright pixels arising from coherent path interference, which behaves differently across frequency. We therefore fit the derotation angle independently in the 16 bands.


Practically, the amount of derotation is measured by fitting the derotation angle $\phi_k$ such that the dominant power (the main arc) of the derotated map $\text{CCWF}e^{-i\phi_k}$ is set to purely real for individual bands $k$. In other words, this is minimizing the sum of the imaginary part of $\text{CCWF}e^{-i\phi_k}$ while maximizing its real part with an optimized value $\phi_k$. Physically, this act is setting the line-of-sight component as the reference phase, since the main arc arises from the coherent interference between the line-of-sight signal and other scattered propagation paths. Its high S/N and dominance in power provide a stable phase reference across frequencies. Therefore, the derotated phase represents the differential propagation effect relative to the line of sight, providing robustness to measure the phase for the 1-ms feature. In Figure 6 of the main text, we show the amount of derotation for the wavefields of the Arecibo and Green Bank Telescope.

\section{Derivation of error propagation}
The measurement of phase is taken to be
\begin{equation}
    \bar\phi=\arctan\left(\frac{\bar I}{\bar R}\right)
\end{equation}
where $\bar R,\bar I$ are the weighted means of the real and imaginary parts of the derotated cross-conjugate wavefield (CCWF) that captures the phase difference of RCP and LCP.

Given the uncertainty in the real and imaginary parts to be $\sigma_{\bar{R}}$ and $\sigma_{\bar I}$, which is obtained by taking 8 random noisy cutout regions from the CCWF, applying the same weighting (after transforming into $\theta-\theta$ space) and get their standard deviations. By taking first order approximation and assuming $\sigma_{\bar{R}}$ and $\sigma_{\bar{I}}$ are independent, the error propagates to $\sigma_{\bar\phi}$ with
\begin{align}
    \sigma_{\bar \phi} &= \sqrt{\left(\frac{\partial \bar\phi}{\partial\bar R}\right)^2\sigma_{\bar{R}}^2 +\left(\frac{\partial \bar\phi}{\partial\bar I}\right)^2\sigma_{\bar I}^2}\\
    &= \sqrt{\left(\frac{-\bar I/\bar R^2}{1+(\bar I/\bar R)^2}\right)\sigma_{\bar R}^2+\left(\frac{1/\bar R}{1+(\bar I/\bar R)^2}\right)\sigma_{\bar I}^2}\\
    &=\frac{|\bar R|}{\bar R^2 + \bar I^2}\sqrt{\left(\frac{\bar I}{\bar R}\right)^2\sigma_{\bar R}^2+\sigma_{\bar I}^2}\ .
\end{align}

\end{document}